\documentclass[review]{elsarticle}

\usepackage{lineno,hyperref}
\usepackage{gensymb}
\usepackage{graphicx}
\usepackage{mathtools}
\usepackage{epstopdf}
\usepackage{float}
\modulolinenumbers[5]

\journal{Icarus}

\begin{document}

\begin{frontmatter}

\title{The compositional diversity of non-Vesta basaltic asteroids}

\author[Lowell,CfA]{Thomas B. Leith}
\author[Lowell]{Nicholas A. Moskovitz}
\author[TCU]{Rhiannon G. Mayne}
\author[MIT]{Francesca E. DeMeo}
\author[USGS]{Driss Takir}
\author[Lowell,MIT]{Brian J. Burt}
\author[MIT]{Richard P. Binzel}
\author[MIT]{Dimitra Pefkou}

\address[Lowell]{Lowell Observatory, Flagstaff, AZ, 86001, USA}
\address[CfA]{Harvard-Smithsonian Center for Astrophysics, Cambridge, MA, 02138, USA}
\address[TCU]{Monnig Meteorite Collection, Texas Christian University, Fort Worth, TX, 76129, USA}
\address[MIT]{Massachusetts Institute of Technology, Cambridge, MA, 02139, USA}
\address[USGS]{Astrogeology Science Center, United States Geological Survey, Flagstaff, AZ, 86001, USA}

\begin{abstract}
We present near-infrared (0.78-2.45 $\mu$m) reflectance spectra for nine middle and outer main belt ($a>2.5$ AU) basaltic asteroids. Three of these objects are spectrally distinct from all classifications in the Bus-DeMeo system and could represent spectral end members in the existing taxonomy or be representatives of a new spectral type. The remainder of the sample are classified as V- or R-type. All of these asteroids are dynamically detached from the Vesta collisional family, but are too small to be intact differentiated parent bodies, implying that they originated from differentiated planetesimals which have since been destroyed or ejected from the solar system. The 1- and 2-$\mu$m band centers of all objects, determined using the Modified Gaussian Model (MGM), were compared to those of 47 Vestoids and fifteen HED meteorites of known composition. The HEDs enabled us to determine formulas relating Band 1 and Band 2 centers to pyroxene ferrosilite (Fs) compositions. Using these formulas we present the most comprehensive compositional analysis to date of middle and outer belt basaltic asteroids. We also conduct a careful error analysis of the MGM-derived band centers for implementation in future analyses. The six outer belt V- and R-type asteroids show more dispersion in parameter space than the Vestoids, reflecting greater compositional diversity than Vesta and its associated bodies. The objects analyzed have Fs numbers which are, on average, between five and ten molar percent lower than those of the Vestoids; however, identification and compositional analysis of additional outer belt basaltic asteroids would help to confirm or refute this result. Given the gradient in oxidation state which existed within the solar nebula, these results tentatively suggest that these objects formed at either a different time or location than 4 Vesta.
\end{abstract}

\end{frontmatter}

\section{Introduction} \label{Intro}

Basaltic asteroids provide us with a rare window into the earliest periods in the formation of the solar system. These objects, which are a relative rarity in the main belt, are thought to form only in the mantle of planetesimals which have undergone the process of metal-silicate differentiation (Gaffey et al., 1993, 2002). Models suggest that these planetesimals were tens to many hundreds of km in size (Hevey and Sanders, 2006; Moskovitz and Gaidos, 2011; Moskovitz and Walker, 2011; Neumann et al., 2012). Compositional analysis of iron meteorites suggests that at least 60 such objects were once present in the inner solar system (Chabot and Haack, 2006; Burbine et al., 2002). Nearly all basaltic asteroids in the main belt are members of the Vesta family, and most are thought to be ejecta from the impact that created Rheasilvia, an impact basin which covers Vesta's southern hemisphere and is one of the largest such features in the solar system (Marazari et al., 1996; McSween et al., 2013). The HED (Howardite, Eucrite, Diogenite) meteorites are thought to also originate from within the Vestoid family, making them a rare example of meteorites which can be tied to a specific differentiated parent body. Eucrites are thought to be pieces of Vesta's crust, Diogenites are thought to originate within Vesta's upper mantle, and Howardites are a brecciated mix of both (Burbine, et al., 2001).

\begin{figure}[H]
\centering
\includegraphics[scale=.70]{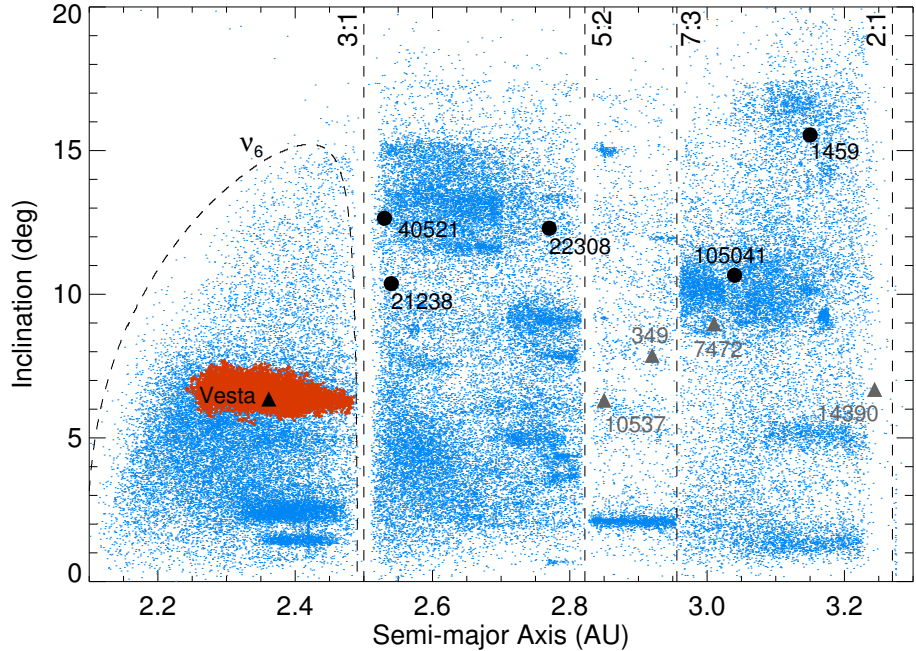}
\caption{Dynamical map of the main belt, showing Vesta (black triangle), the Vestoids (red dots), our V-type targets (black circles), and non-V-types discussed in this work (gray triangles). Other asteroids are shown as blue dots. The dashed lines show the location of major mean motion and secular resonances. It is dynamically unlikely that any of our targets are derived from Vesta.}
\label{fig:dynamics}
\end{figure}

A few basaltic asteroids which are not associated with Vesta's dynamical family have been discovered in recent years. These objects lie in the middle and outer main belt across Jupiter's 3:1 mean motion resonance from Vesta (see Figure \ref{fig:dynamics}). Roig et al. (2008) show that there is a less than 1 percent chance of a 5-km V-type asteroid being ejected from Vesta and crossing the 3:1 Kirkwood gap. Several outer belt basaltic asteroids, such as (1459) Magnya, are also simply too large to have plausibly been ejected from Vesta (Ieva et al., 2015). Furthermore, the Dawn spacecraft has provided sufficiently high spatial resolution of Vesta's surface spectral features to conclude that at least two V-types beyond the 3:1 Kirkwood gap could not have originated from the Rheasilvia impact basin (Ieva et al., 2015). Thus, they are most likely fragments of differentiated parent bodies which have since been destroyed or ejected from the solar system  (Lazzaro et al., 2000; Hammergren et al., 2006; Roig et al., 2008; Huaman et al., 2014). 

To further understand these objects, along with how and where they may have formed, we have analyzed near-infrared reflectance spectra of nine basaltic asteroids located in the main belt between $a=2.5$ AU and $a=3.3$ AU. We then compared the spectra of these asteroids to those of 47 Vestoid asteroids and fifteen HED meteorites of known composition. Eight additional achondrite, olivine-rich meteorites from the Acapulcoite and Lodranite suites were analyzed to serve as references for objects that display 1- and 2-$\mu$m absorption bands but do not have HED-like compositions. While previous studies have examined several of these asteroids before, there has yet to be a comprehensive analysis of a majority of the known middle and outer belt basaltic asteroids. Moskovitz et al. (2008c), Hardersen et al. (2004), and Roig et al. (2008) performed focused studies of single objects, while Ieva et al. (2015) analyzed only two outer belt V-types---(1459) Magnya and (21238) Panarea, both of which are discussed in this work. Our study examines a larger sample of basaltic bodies past the 3:1 Kirkwood gap in order to broadly assess these objects' composition and make inferences about where this population could have originated. The location of these objects' orbits within the main belt is shown in Figure \ref{fig:dynamics}.

Throughout this work we use several terms. V-type asteroids, the category to which most of our target objects belong, are a spectral type within the Bus-DeMeo asteroid taxonomic system (Bus and Binzel, 2002b; DeMeo et al., 2009), distinguished by deep absorption bands near 1 and 2 $\mu$m which are characteristic of a pyroxene composition. One of our target asteroids, (349) Dembowska, is not a V-type, but rather is the only known R-type according to the Bus-DeMeo taxonomic system (Bus and Binzel, 2002b; DeMeo et al. 2009). The R-type spectrum also shows significant absorption features near 1 and 2 $\mu$m, but has a slightly flatter spectrum and a more complex 1-$\mu$m absorption feature, likely due to the presence of olivine (Gaffey et al., 1993). The Vestoid family is a dynamical and compositional grouping of V-type asteroids associated with 4 Vesta, the second-most massive body in the main belt and the only asteroid believed to be a largely intact differentiated planetesimal. Other types of planetary differentiation are known to occur, such as ice-silicate differentiation---however, when used in this work, the term differentiation refers exclusively to metal-silicate differentiation.

The process of metal-silicate differentiation was most likely driven by radioactive decay of short-lived isotopes such as $^{26}$Al and $^{60}$Fe (Grimm and McSween, 1993; Goswami and Vanhala, 2000; Tachibana and Huss, 2003), which have half-lives of 0.73 and 1.5 Myr, respectively. This places constraints on both the time and location of these objects' formation, as they must have undergone differentiation within a few half-lives and could not have formed in the gas giant region, where they would have accreted too slowly to undergo melting (Moskovitz et al., 2008a). Bottke et al. (2006) propose that metal-silicate differentiation occurred in the inner solar system, interior to $a=2.0$ AU. This would indicate that Vesta, the Vestoids, and the other main belt basaltic asteroids which are the focus of this study may have formed at a considerably smaller heliocentric distance than their current locations and were later scattered outward. 

In order to further explore the possibility of this scenario, we analyzed the pyroxene mineralogy of these asteroids. Pyroxenes have varying abundances of iron, magnesium, and calcium, the molar percentages of which are reflected in their respective Ferrosilite (Fs), Enstatite (En), and Wollastonite (Wo) numbers (e.g. Mayne et al., 2010). In this work, we focused on determining the Fs number of each of our target asteroids. In addition to scaling with the bulk iron concentration in which it formed, a given pyroxene's Fs number is strongly influenced by the oxidation state of its formation environment, since oxidizing conditions tend to produce more iron-rich minerals. Given that an oxidation gradient existed within the solar nebula, comparing two objects' Fs numbers can thus shed light on where they formed relative to one another (Rubin and Wasson, 1995). 

To take advantage of this phenomenon, we derived formulas relating the centers of pyroxene absorption bands from HED meteorite spectra taken from the RELAB (Reflectance Experiment Laboratory) database at Brown University. We then used them to calculate the Fs content of our target objects. In doing so we aim to provide observational constraints on the Bottke et al. (2006) model of scattering in the young solar system.

The remainder of this paper is organized as follows: in Section 2 we discuss how and when our asteroid observations were obtained. In Section 3 we fit observational spectra of our target objects, Vestoids, and HED meteorites using the Modified Gaussian Model (Sunshine et al. 1990; Sunshine et al. 1993) and relate spectral properties to composition. In Section 4 we discuss the spectral classification of our objects, and interpret the results in light of the Bottke et al. (2006) model of scattering and the ``Grand Tack" planetary migration model put forward by Walsh et al. (2012).

\section{Observations and reduction} \label{Obs}

\subsection{Asteroid observations}

Much of the data for this work were drawn from published, archival sources. For objects with little or no existing spectral data, we obtained new near-infrared (0.8-2.5 $\mu$m) spectra with two instruments: SpeX at NASA's Infrared Telescope Facility (IRTF) on Mauna Kea (Rayner et al., 2005) and the Folded-port Infrared Echellette (FIRE) spectrograph on the Magellan Baade 6.5-m at Las Campanas in Chile (Simcoe et al., 2008). The sources of visible (where applicable) and near-IR spectra for each of our target objects are shown in Table \ref{tab:ObsSource}. In all cases except the asteroids (349) Dembowska and (1459) Magnya, our middle and outer belt targets were originally identified as candidate basaltic asteroids based on their photometric colors in the Sloan Digital Sky Survey Moving Object Catalog (SDSS MOC; Ivezic et al. 2001). The {\it ugriz} photometric band passes employed by the SDSS MOC enable coarse assignment of asteroid spectral types and are particularly well suited at distinguishing basaltic asteroids with deep 1 $\mu$m absorption features (Rig and Gil-Hutton, 2006; Moskovitz et al. 2008b; Solontoi et al. 2012). The only two targets not selected for follow-up based on SDSS colors were (349) Dembowska, long known to have unusual spectral characteristics (Feierberg et al. 1980), and (1459) Magnya, discovered as an outer belt V-type as part of the larger S3OS2 survey (Lazzaro et al. 2000).

\begin{table}[H]
\tiny
\centering
\begin{tabular}{| l l l |}
\hline
\multicolumn{3}{ | c | }{Sources of observational data}\\
\hline
Object & Visible data & NIR data\\
\hline
(349) Dembowska & Bus et al., 2002 & DeMeo et al. 2009\\
(1459) Magnya & Lazzaro et al. 2000 & Hardersen et al. 2004\\
(7472) Kumakiri & No visible data & IRTF, Sept 6, 2010\\
(10537) 1991 RY$_{16}$ & Moskovitz et al. 2008b & Moskovitz et al. 2008b\\
(14390) 1990 QP$_{19}$ & No visible data & IRTF, July 11, 2013\\
(21238) Panarea (1) & Roig et al. 2008 & FIRE, Aug 24, 2010\\
(21238) Panarea (2) & Roig et al. 2008 & IRTF,  Sept 8, 2015\\
(22308) 1990 UO$_4$ & No visible data & FIRE, July 3, 2012\\
(40521) 1999 RL$_{95}$ & No visible data & IRTF, Sept 14, 2015\\
(105041) 2000 KO$_{41}$ & Solontoi et al. 2012 & FIRE, Oct 24, 2012\\
\hline
\end{tabular}
\caption{Sources of all observational data for target asteroids. Spectra of these objects are shown in Figure \ref{fig:outliers}.}
\label{tab:ObsSource}
\end{table}

Acquisition of data from these two instruments followed analogous procedures. For both the slit mask was oriented along the parallactic angle at the start of each observation to minimize the effects of atmospheric dispersion. Solar analogs were observed to correct for telluric absorption and to remove the solar spectrum from the measured reflectance. Objects and solar analogs were observed near the meridian and at similar airmass. Individual exposures for the asteroids were held between 120-180s to avoid saturation of telluric emission features and, for FIRE, to avoid saturated thermal emission from the instrument and telescope at wavelengths longer than about 2.2 $\mu$m. Exposures were obtained in standard ABBA nod sequences with the target offset by several arc seconds along the slit at the two nod positions.

FIRE was operated in its high-throughput prism mode with a 0.8 $\times$ 50'' slit. These settings produced single-order spectra at a resolution of approximately 400 from 0.8 to 2.45  $\mu$m. At the IRTF, SpeX was configured in its low resolution (R = 250) prism mode with a 0.8" slit for wavelength coverage from 0.8 to 2.5 $\mu$m.  Reduction of the SpeX data followed DeMeo et al. (2009). Reduction of the FIRE data employed an IDL package designed for the instrument and based on the Spextool pipeline (Cushing et al., 2004) using procedures detailed in Binzel et al. (2015). 

\begin{figure}[H]
\centering
\includegraphics[scale=.63]{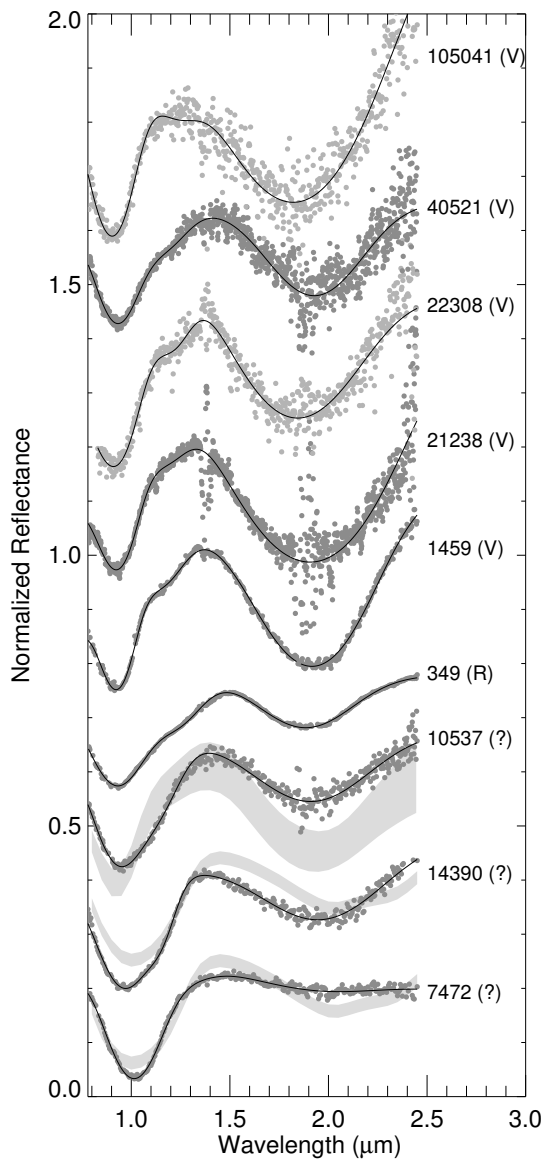}
\caption{Normalized and vertically offset spectra and MGM fits to each of the middle and outer belt objects discussed in this work. Data points are plotted as solid circles, and the MGM fit as a solid black line. The numerical designation of each object is listed at right with its spectral type, where known. Objects with an ambiguous spectral classification---i.e. not consistent with any known types---are denoted with a ``?''. These three spectra are  plotted with the closest available type to which they could be assigned (from top to bottom: V, O, and O).}
\label{fig:outliers}
\end{figure}

Spectra for all of our outer belt target objects are shown in Figure \ref{fig:outliers}. In addition to these observations, we utilized spectral data on 4 Vesta from Gaffey (1997) and a large sample of Vestoid spectra previously compiled in Moskovitz et al. (2010). We calculated signal-to-noise for each band center independently by taking the inverse of the standard deviation of the residuals to our fits (see section \ref{Bands}) in the region of each band center. We defined this region for Band 1 as 0.85-1.05 $\mu$m, and for Band 2 as 1.85-2.05 $\mu$m. Observational circumstances for each of our targets are summarized in Table \ref{tab:astObs}.

\subsection{Meteorite spectra}

We compared our asteroid data to reflectance spectra of HED meteorites taken using the near-IR bidirectional spectrometer at at the NASA/Keck Reflectance Experiment Laboratory at Brown University (Pieters, 1983). In addition, we analyzed reflectance spectra of Acapulcoite and Lodranite meteorites taken at the same facility. The HED meteorites provided spectra of Vestoid analogs with known, laboratory-determined compositions. The Acapulcoites and Lodranites, meanwhile, served as comparators with similar spectral features (i.e. 1- and 2-$\mu$m absorption bands). These specific groups were of particular interest as they are known to be more olivine-rich than the Vestoids and HED meteorites (e.g. McCoy et al., 1996) and thus can provide insight on the limitations of our methodology for non-Vesta compositions. The process of compositional validation performed using these meteorite spectra is outlined in sections \ref{Composition} and \ref{ComparisonMets}.

For each sample we analyzed spectral data taken from observations of the smallest grain size available. Specific grain sizes for each meteorite measured are available in Table \ref{Table:hedMGM}, along with Fs numbers drawn from either Mayne et al. (2011) or Burbine et al. (2009) as indicated.

\begin{table}[H]
\tiny
\centering
\begin{tabular}{| l l l l l l l l |}
\hline
\multicolumn{8}{ | c | }{Vesta and Vestoids}\\
\hline
Object & UT Date & $r (AU)$ & $\alpha (\degree)$ & Mag. & $t_{exp}$ (min) & S/N (B1) & S/N (B2) \\
\hline
(4) Vesta & Feb 18-20 , 1981 & 2.38 & 4 & 6.1 & - & 523 & 442 \\
(809) Lundia & Aug 26 , 2008 & 1.93 & 23.9 & 14.6 & 8 & 560 & 168 \\
(956) Elisa & Jul 5 , 2008 & 1.85 & 16.6 & 14.6 & 16 & 637 & 209 \\
(1468) Zomba & Sep 30 , 2003 & 1.6 & 38.3 & 16.5 & 32 & 534 & 226 \\
(1929) Kollaa & Feb 19 , 2001 & 2.21 & 14.4 & 15.3 & 24 & 296 & 198 \\
(2045) Peking & Aug 26 , 2008 & 2.5 & 16.9 & 16.2 & 87 & 531 & 154 \\
(2371) Dimitrov & Aug 1 , 2009 & 2.47 & 19.6 & 16.7 & 40 & 260 & 42 \\
(2442) Corbett & Sep 15 , 2002 & 2.25 & 7.8 & 15.6 & 12 & 371 & 146 \\
(2511) Patterson & May 7 , 2004 & 2.23 & 21.2 & 16.1 & 32 & 325 & 197 \\
(2566) Kirghizia & May 8 , 2002 & 2.41 & 9.8 & 16 & 32 & 300 & 203 \\
(2579) Spartacus & Oct 10 , 2000 & 2.22 & 18.9 & 16.4 & 28 & 152 & 72 \\
(2653) Principia & Nov 26 , 2002 & 2.53 & 17.8 & 16.3 & 48 & 504 & 281 \\
(2763) Jeans & Jun 26 , 2004 & 2.24 & 12.6 & 15.7 & 32 & 279 & 358 \\
(2795) Lepage & Apr 9 , 2005 & 2.26 & 4.7 & 15.9 & 24 & 354 & 423 \\
(2823) van der Laan & Nov 22 , 2005 & 2.2 & 16.1 & 16.4 & 8 & 302 & 38 \\
(2851) Harbin & Aug 24 , 2001 & 2.42 & 8.7 & 15.6 & 28 & 437 & 312 \\
(2912) Lapalma & Feb 20 , 2001 & 2.14 & 8.7 & 15.3 & 32 & 354 & 225 \\
(3155) Lee & Jun 22 , 2001 & 2.56 & 8.4 & 16.2 & 28 & 344 & 129 \\
(3344) Modena & Sep 4 , 2005 & 2.17 & 17.9 & 16.1 & 32 & 456 & 182 \\
(3657) Ermolova & Aug 1 , 2009 & 2.18 & 25.4 & 16.6 & 40 & 380 & 72 \\
(3703) Volkonskaya & Jun 3 , 2006 & 2.25 & 9.8 & 17.2 & 48 & 317 & 114 \\
(3782) Celle & Nov 26 , 2002 & 2.6 & 17.1 & 16.8 & 32 & 295 & 226 \\
(4038) Kristina & Oct 28 , 2002 & 2.08 & 10.3 & 15.8 & 48 & 594 & 186 \\
(4188) Kitezh & Aug 14 , 2001 & 2.1 & 10.8 & 15.3 & 28 & 356 & 256 \\
(4215) Kamo & Nov 11 , 2002 & 2.52 & 17.3 & 16.5 & 64 & 385 & 140 \\
(4796) Lewis & Jan 9 , 2009 & 2.46 & 13.9 & 17 & 33 & 117 & 55 \\
(5481) Kiuchi & Sep 5 , 2005 & 2.37 & 10.5 & 16.3 & 54 & 327 & 96 \\
(5498) Gustafsson & Aug 26 , 2008 & 1.92 & 7.8 & 16.3 & 93 & 861 & 212 \\
(7800) Zhongkeyuan & Jan 9 , 2009 & 2.51 & 2.8 & 17.1 & 27 & 190 & 31 \\
(9481) Menchu & Aug 26 , 2008 & 2.48 & 6.1 & 17.3 & 40 & 146 & 40 \\
(9553) Colas & Jan 8 , 2009 & 1.99 & 20.3 & 17.4 & 120 & 359 & 124 \\
(16416) 1987 SM3 & Nov 23 , 2007 & 2.4 & 6 & 16.9 & 64 & 114 & 51 \\
(26886) 1994 TJ2 & Jul 5 , 2008 & 2.09 & 10.9 & 17.2 & 40 & 263 & 40 \\
(27343) 2000 CT102 & Aug 26 , 2008 & 1.94 & 11.7 & 16.6 & 48 & 323 & 82 \\
(33881) 2000 JK66 & Nov 23 , 2007 & 1.77 & 23.3 & 16.6 & 48 & 316 & 81 \\
(36412) 2000 OP49 & Nov 23 , 2007 & 2.12 & 11.8 & 16.9 & 40 & 123 & 50 \\
(38070) 1999 GG2 & Oct 5 , 2006 & 1.9 & 4.1 & 16.9 & 32 & 304 & 63 \\
(50098) 2000 AG98 & Sep 4 , 2005 & 1.94 & 11.4 & 15.7 & 24 & 473 & 240 \\
(97276) 1999 XC143Â & Nov 23 , 2007 & 2.05 & 11.4 & 15.7 & 24 & 182 & 80 \\
\hline   
\multicolumn{8}{ | c | }{Outer Belt V- and R-Types}\\               
\hline 
(349) Dembowska (1) & Jun 21 , 2001 & 2.956 & 17.9 & 10.5 & 8 & - & - \\
(349) Dembowska (2) & November 26 , 2001 & 2.71 & 1 & 10.5 & 11 & 320 & 389 \\
(1459) Magnya & Mar 23 , 2003 & 3.851 & 17.3 & 16.3 & 40 & 197 & 167 \\
(21238) Panarea (1) & Jul 20 , 2006 & 2.8 & 5.5 & 17.1 & 14 & 97 & 32 \\
(21238) Panarea (2) & Sept 8 , 2015 & 1.84 & 21.6 & 17.5 & 56 & 168 & 14 \\
(22308) 1990 UO$_4$ & Jul 3 , 2012 & 2.55 & 15 & 17.8 & 30 & 77 & 35 \\
(40521) 1999 RL$_{95}$ & Sept 14 , 2015 & 1.41 & 3.1 & 17.6 & 24 & 152 & 23 \\
(105041) 2000 KO$_{41}$ & Oct 24 , 2012 & 2.89 & 7.3 & 18.6 & 48 & 89 & 25 \\
\hline   
\multicolumn{8}{ | c | }{Spectrally Anomalous Targets (See Section \ref{NewType})}\\              
\hline           
(7472) Kumakiri & Sep 6-7 , 2010 & 3.33 & 6 & 17 & 64 & 241 & 143 \\
(10537) 1991 RY$_{16}$ & Jan 30 , 2008 & 2.85 & 3.7 & 16.4 & 137 & 223 & 50 \\
(14390) 1990 QP$_{19}$ & Jul 11 , 2008 & 2.9 & 19.7 & 18 & 24 & 301 & 107 \\
\hline
\end{tabular}
\caption{Observational data for main belt basaltic asteroids. The columns in this table are: object designation, UT date of observation, heliocentric distance at the time of observation in AU ($r$), phase angle of the target in degrees ($\alpha$), V-band magnitude of the target (V, retrieved from Moskovitz et al. 2010), the net exposure time in minutes, and S/N in the region of each band center.}
\label{tab:astObs}
\end{table}
\section{MGM band fitting} \label{MGMsec}

\begin{figure}[H]
\centering
\includegraphics[scale=.45]{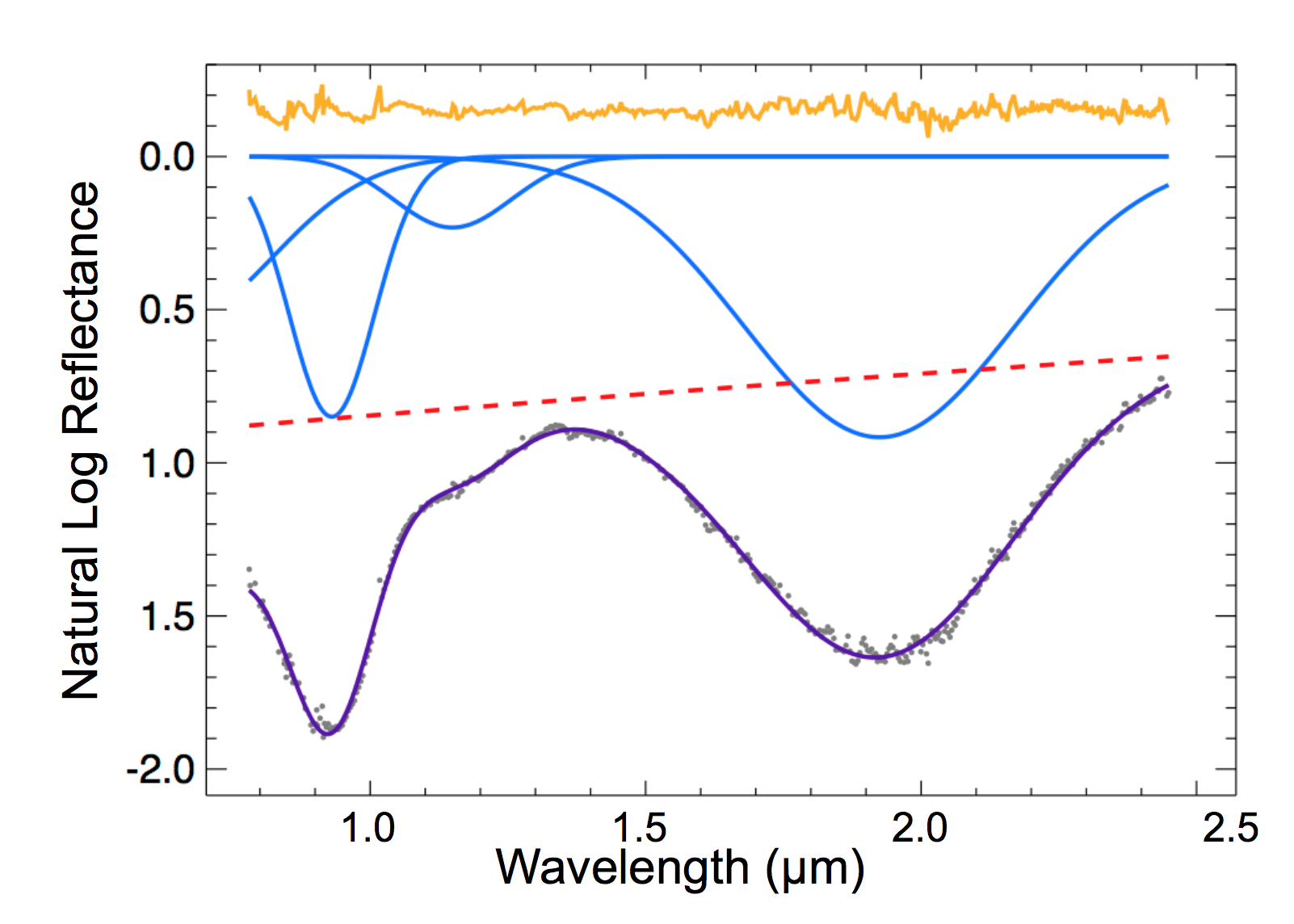}
\caption{MGM fit to the spectrum of the outer belt V-type (1459) Magnya. The data are represented by gray dots, and the continuum by a red dashed line. The modeled bands are in blue, the fit in purple, and the residuals in orange. MGM allows for highly accurate fits to complex features like the overlapping bands at 0.9 and 1.2 $\mu$m and is effective at isolating the characteristics of individual bands.}
\label{fig:mgmfit}
\end{figure}

\subsection{Band analysis} \label{Bands}

1-$\mu$m and 2-$\mu$m band centers for all asteroid and meteorite spectra were determined using the Modified Gaussian Model (MGM) (Sunshine et al. 1990, Sunshine et al. 1993). MGM uses a flexible continuum and a series of modified Gaussians representing absorption bands to fit a spectrum. Parameters representing the offset, slope, and curvature of the continuum, as well as the number, location, width, and depth of bands are supplied by the user, then adjusted by MGM within a user-defined range until it has achieved an optimal fit. MGM is thus able to accurately fit spectra and determine individual band parameters even in the presence of overlapping bands, which is particularly important when analyzing pyroxene spectra featuring absorption bands centered near 1 $\mu$m and 1.2 $\mu$m (Mayne et al., 2010). 

Previous research using MGM to analyze basaltic asteroid and meteorite spectra has employed a physically-motivated fit in order to yield specific mineralogy (Mayne et al., 2010; Ieva et al., 2015). However, our approach was a purely mathematically-motivated one, similar to that employed by Thomas \& Binzel (2010) and Mayne et al. (2011). We aimed to determine a single set of input parameters which would yield the best possible fit when used with all of our HED meteorite and asteroid spectra while simultaneously minimizing free parameters by using the fewest possible bands. While this approach would not be sufficient for extracting mineralogical information on any single object, applying it to spectra of HED meteorites with known mineralogy and basaltic asteroids yields relative mineralogical insight. These can then be used for comparison across populations of basaltic objects.

To ensure that our MGM fits were well-calibrated to objects with known mineralogy, we began by collecting spectra of meteorites with known bulk pyroxene Fs numbers from the online database curated by the NASA/Keck Reflectance Experiment Laboratory (RELAB) at Brown University (Pieters, 1983). These spectra were then truncated to the NIR region of 0.78-2.45 $\mu$m in order to match the minimum spectral range of our outer belt asteroid spectra. We then attempted MGM fits to our HED meteorite spectra using a variety of input parameters fitting 3, 4, and 5 bands with flat, sloped, and nonlinear continuums. The success of each set of parameters was judged by both the quality of the fit as measured by the overall RMS error output by MGM and by how closely the resulting band centers matched those found in Mayne et al. (2011). RELAB does not provide error bars on its spectra. However, we were able to estimate a wavelength-averaged signal-to-noise (S/N) of each spectrum as the mean reflectance value divided by the standard deviation of the RMS residuals of the fit, which we then mapped to band center errors using the methodology outlined in section \ref{error}.

We ultimately found that a set of parameters modeling four bands and a sloped continuum yielded the best fit while minimizing free parameters. However, in the case of three Eucrite meteorites---ALHA81001, BTN00300, and MAC02522---the output band centers varied too significantly from those measured in Mayne et al. (2011), and so these objects were excluded from our sample. 

Our exact set of input parameters was further refined by using MGM to fit spectra of Vesta, numerous V-type asteroids, and our sample of basaltic outer belt asteroids. For several objects, MGM fit an unrealistically wide 1.2-$\mu$m band, which resulted in the 1-$\mu$m band falling well below where it could plausibly be located. As a result, the maximum width of the 1.2-$\mu$m band was restricted significantly more than that of other bands. Ultimately we were able to determine a single set of input parameters which could be utilized for MGM fits to all of our meteorite and asteroid spectra, with the lone exception of the outer belt asteroid (14390) 1990 QP$_{19}$. This object, along with asteroids (7572) Kumakiri and (10537) 1991 RY16, may represent a new spectral type of asteroid and is discussed at greater length in section \ref{NewType}. Our final set of input parameters is shown in Table \ref{tab:fitparam}, and a sample MGM fit (to (1459) Magnya) is shown in figure \ref{fig:mgmfit}.

\begin{table}[H]
\tiny
\centering
\begin{tabular}{| l l l l l l |}
\hline
Object & RELAB  & Grain Size  & Measured Fs & B1 Center & B2 Center \\
           & designation &  ($\mu$m)      &  Number        & ($\mu$m) & ($\mu$m)\\
\hline
\multicolumn{6}{| l |}{Diogenites}\\
EETA79002	&	MB-TXH-067-A	&	0-25	&	22.0$^a$	&	0.917	$\pm$	0.003	&	1.889	$\pm$	0.004	\\
GRO95555	&	MP-TXH-068-A	&	0-25	&	25.0$^a$	&	0.921	$\pm$	0.003	&	1.903	$\pm$	0.004	\\
Johnstown	&	MB-TXH-095-A	&	0-25	&	23.5$^b$	&	0.916	$\pm$	0.002	&	1.881	$\pm$	0.003	\\
LAP91900	&	MP-TXH-077-A	&	0-25	&	23.0$^a$	&	0.919	$\pm$	0.003	&	1.906	$\pm$	0.004	\\
Tatahouine	&	MP-TXH-088-A	&	0-25	&	23.0$^b$	&	0.918	$\pm$	0.002	&	1.893	$\pm$	0.003	\\
\multicolumn{6}{| l |}{ }\\															
\multicolumn{6}{| l |}{Howardites}\\															
EET87503	&	MB-TXH-068-AP	&	0-25	&	36.5$^a$	&	0.928	$\pm$	0.003	&	1.954	$\pm$	0.004	\\
QUE94200	&	MP-TXH-069-A	&	0-25	&	30.5$^a$	&	0.921	$\pm$	0.003	&	1.921	$\pm$	0.003	\\
Y791573	&	MP-TXH-099-A	&	0-25	&	-	&	0.927	$\pm$	0.004	&	1.963	$\pm$	0.005	\\
\multicolumn{6}{| l |}{ }\\															
\multicolumn{6}{| l |}{Eucrites}\\															
Chervony Kut	&	MT-HYM-035	&	0-38	&	41.9$^a$	&	0.934	$\pm$	0.003	&	2.005	$\pm$	0.004	\\
EET87520	&	MT-HYM-029	&	0-45	&	39.7$^a$	&	0.946	$\pm$	0.003	&	2.016	$\pm$	0.004	\\
GRA98098	&	MT-HYM-034	&	0-38	&	51.2$^a$	&	0.937	$\pm$	0.003	&	1.996	$\pm$	0.004	\\
Ibitira	&	MP-TXH-054-A	&	0-25	&	42.0$^a$	&	0.938	$\pm$	0.002	&	1.990	$\pm$	0.003	\\
MET01081	&	MT-HYM-033	&	0-45	&	44.6$^a$	&	0.934	$\pm$	0.003	&	1.982	$\pm$	0.004	\\
Moore County	&	MP-TXH-086-A	&	0-25	&	37.3$^a$	&	0.935	$\pm$	0.003	&	1.982	$\pm$	0.004	\\
PCA91078	&	MT-HYM-031	&	0-45	&	40.2$^a$	&	0.953	$\pm$	0.005	&	2.021	$\pm$	0.005	\\
Serra de Mage	&	MP-TXH-092-A	&	0-25	&	36.2$^a$	&	0.927	$\pm$	0.003	&	1.963	$\pm$	0.004	\\
\hline															
\multicolumn{6}{| l |}{Acapulcoites}\\															
Acapulco	&	MT-CMP-001	&	0-25	&	8.8$^c$	&	0.933	$\pm$	0.005	&	1.809	$\pm$	0.006	\\
ALHA81187	&	TB-TJM-036	&	0-25	&	5.1$^c$	&	0.907	$\pm$	0.003	&	1.862	$\pm$	0.004	\\
ALHA81261	&	TB-TJM-039	&	0-25	&	7.5$^c$	&	0.935	$\pm$	0.005	&	1.845	$\pm$	0.005	\\
\multicolumn{6}{| l |}{ }\\															
\multicolumn{6}{| l |}{Lodranites}\\															
Lodran	&	TB-TJM-034	&	0-25	&	10.1$^e$	&	0.918	$\pm$	0.004	&	1.863	$\pm$	0.004	\\
GRA95209	&	TB-TJM-037	&	0-25	&	2.4-4.7$^d$	&	0.916	$\pm$	0.004	&	1.877	$\pm$	0.004	\\
MAC88177	&	MB-CMP-026	&	0-25	&	9.1$^e$	&	0.948	$\pm$	0.005	&	1.895	$\pm$	0.005	\\
Y74357	&	MB-TXH-037	&	0-25	&	10.1$^e$	&	0.945	$\pm$	0.003	&	1.881	$\pm$	0.003	\\
Y791491	&	MB-TXH-038	&	0-25	&	8.5$^e$	&	0.932	$\pm$	0.003	&	1.864	$\pm$	0.004	\\
\hline
$^a$Mayne et al. (2011) & & & & &\\
$^b$Burbine et al. (2009)& & & & &\\
$^c$McCoy et al. (1996)& & & & &\\
$^d$Floss (1991)& & & & &\\
$^e$McCoy et al. (1997)& & & & &\\
\hline
\end{tabular}
\caption{MGM fit results for HED meteorites and Acapulcoite and Lodranite meteorites modeled for comparison purposes. The columns in this table are: object designation, RELAB database designation, B1 center, B2 center, Fs number, grain size. Error on meteorite band centers is determined using the methodology outlined in section \ref{error}.}
\label{Table:hedMGM}
\end{table}

\begin{table}[H]
\tiny
\centering
\begin{tabular}{| l l l l |}
\hline
Continuum offset & & Continuum slope & \\
5.000 $\times 10^{-1}$ $\pm$ 0.1 & & -3.986 $\times 10^{-5}$ $\pm$ 5.0 $\times 10^{-5}$ & \\
 & & & \\
Band designation & Band center ($\mu$m) & Band FWHM ($\mu$m) & Band strength \\
A & 0.450 $\pm$ 0.075 & 0.397 $\pm$ 0.100 & -0.530 $\pm$ 0.10 \\
Band 1 & 0.935 $\pm$ 0.100 & 0.189 $\pm$ 0.200 & -7.074 $\pm$ 0.50 \\
B & 1.178 $\pm$ 0.100 & 0.250 $\pm$ 0.025 & -0.500 $\pm$ 0.75 \\
Band 2 & 2.010 $\pm$ 0.100 & 0.703 $\pm$ 0.400 & -0.500 $\pm$ 0.50 \\
\hline
\end{tabular}
\caption{Input parameters for MGM fitting. The top section shows continuum parameters and the bottom shows band parameters. The value after the modulus represents the 1-sigma variation allowed for each parameter. Bands designated A and B are located below Band 1 and between Band 1 and 2, respectively.}
\label{tab:fitparam}
\end{table}

\subsection{Band Center Error Analysis} \label{error}

Formal error bars on fitted band centers behave non-linearly as a function of signal-to-noise (S/N). These errors are non-trivial to estimate because typical observational data only contain uncertainty associated with reflectance values whereas the accuracy of band centers is an uncertainty associated with the spectral dispersion axis. Thus the S/N of spectral data cannot be used as a direct indicator of band center uncertainty. 

\begin{figure}[H]
\centering
\includegraphics[scale=.60]{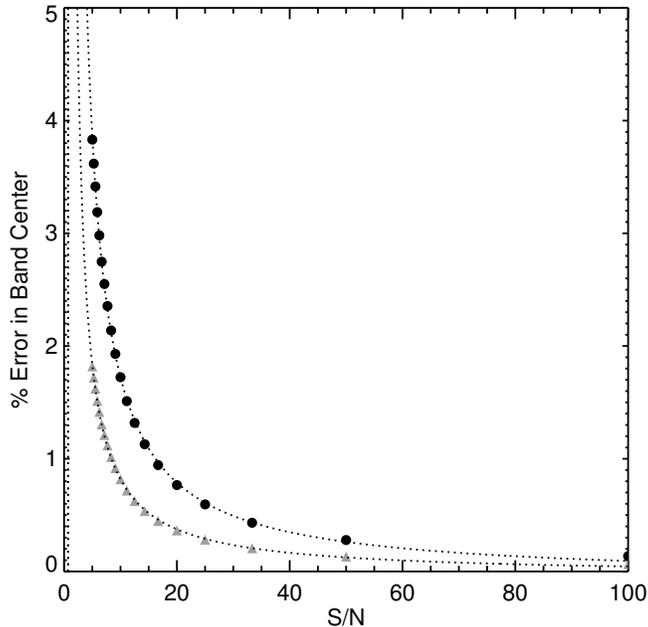}
\caption{$3\sigma$ error in band center location for Band 1 (circles) and Band 2 (triangles) as a function of S/N, calculated according to the process outlined in section \ref{error} for the Howardite Y791573.}
\label{fig:errorplot}
\end{figure}

We performed a series of Monte Carlo experiments to assign formal error bars to MGM-fitted band centers as a function of S/N. We began with RELAB spectra of a Howardite (Y791573), a Eucrite (Ibitira), and a Diogenite (Tatahouine), and then fit their band centers using MGM. These particular samples were chosen because they have exceedingly high S/N ($>>100$) and are well fit with our 4 band MGM model. We then added a prescribed level of random noise to the spectra and re-fit the band centers. This addition of noise and fitting of the spectra was repeated 1000 times. The standard deviation of the band centers from the 1000 trials was then used to assign $3\sigma$ uncertainties to each band center as a function of the prescribed noise level. This process was repeated for a range of typical S/N values from 5 to 100 (Figure \ref{fig:errorplot}). Each of the HED subtypes produced roughly equivalent results.

The relationship between percentage error in band center and the S/N of the spectrum is essential to properly assess uncertainties on the spectral parameters presented here, but can also be a useful tool for future observational planning. For example, if a particular science case requires that a band center be measured to a precision of $\sim1\%$, then the requisite S/N in the region of the band center would be $\sim15$. To better enable the use of these Monte Carlo results for future observations we provide approximate analytical expressions to quantify the percent error in each band as a function of S/N:
\begin{equation}
\% Error_{B1} \sim 16.5 * (S/N)^{-1}
\end{equation}
\begin{equation}
\% Error_{B2} \sim 7.8 * (S/N)^{-1}
\end{equation}

This uncertainty analysis is purely based on statistical (S/N) errors associated with spectral data. This analysis does not apply to systematic errors that could also influence the accuracy of band center fits. For example, outlying points in asteroid reflectance spectra due to incomplete removal of telluric bands around 0.95, 1.1, and 1.9 $\mu$m can skew band centers in ways that are difficult to quantify. Treatment of such systematic effects is beyond the scope of this work.

These band center error estimates are specific to V-type and HED spectra. While these results likely provide a rough estimate on formal band center errors for other compositions (e.g. ordinary chondrites), a similar analysis would be required to properly assess the errors for different spectral types/compositions.

\subsection{Temperature corrections to band centers} \label{TempCorr}

Higher surface temperatures cause a shift in observed band centers to lower wavelengths. Thus, when comparing meteorite laboratory spectra to observational data, it was important to determine the temperature of each asteroid and correct MGM-derived band centers accordingly. Surface temperatures were estimated using the following equation, from Burbine et al. (2009): \begin{equation}\label{eq:tempeq}
T=[(1-A)L_0/16\eta\epsilon\sigma\pi r^2]^{1/4}
\end{equation} Where $A$ is the albedo of the asteroid, $L_0$ is solar luminosity, $\eta$ is the beaming factor (set to unity), $\epsilon$ is the infrared emissivity of the asteroid (set to 0.9), $\sigma$ is the Stefan-Boltzmann constant, and $r$ is the heliocentric distance of the asteroid at the time of its observation. As discussed in Burbine et al. (2009), adjustments on the scale of 0.1 to $\eta$ and $\sigma$ alter the resulting temperature by only $\pm$ 10 K. Albedos for each object were retrieved from the WISE catalog (Mainzer et al. 2011) or IRAS (Tedesco et al., 2002). For objects whose albedo was unavailable, the albedo of 4 Vesta (0.4) was assumed, as it should be representative for V-type asteroids. Albedo errors for all objects had effects on the corrected band centers at or below the order of 0.001\%, and thus were not accounted for. 

Having calculated the surface temperature of each object, we then corrected our MGM-derived band centers using the formulas determined by Burbine et al. (2009). Burbine et al. analyzed two orthopyroxene samples and calculated two formulas for each band correction. We applied these equations to the temperature of each object calculated in Equation \ref{eq:tempeq}, and then averaged the resulting corrections for each pair of equations and applied them to our MGM-derived band centers. The temperature corrections to our asteroid band centers, as well as the temperature-corrected band centers themselves, are shown in Table \ref{tab:astMGM}.

\subsection{Compositional validation} \label{Composition}

\begin{figure}[H]
\centering
\includegraphics[scale=.48]{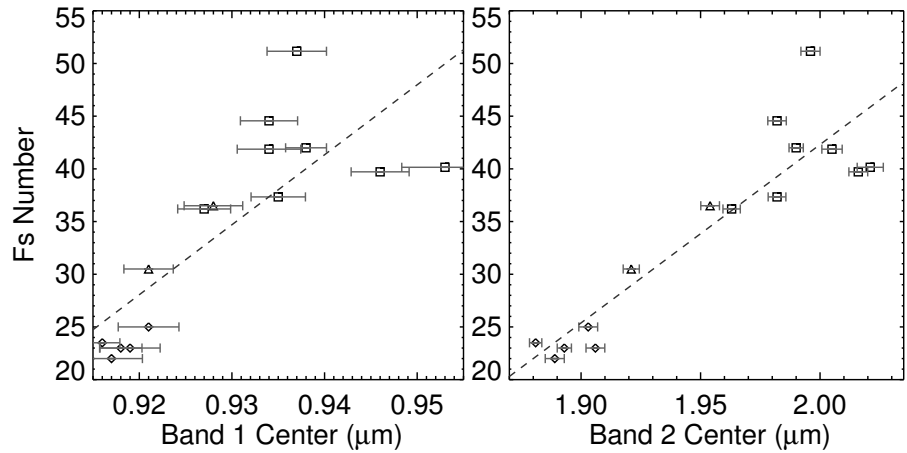}
\caption{Fs number of HED meteorites plotted against Band 1 and Band 2 centers. Diogenites are represented by diamonds, Howardites by triangles, and Eucrites by squares. Each was fitted using the IDL REGRESS function. A clear correlation is evident between band centers and Fs number.}
\label{fig:bandvsfs}
\end{figure}

Our next step was to address how well we could use MGM-derived band centers to constrain the bulk pyroxene Fs number of any given object. We began by calibrating meteorite spectra against their known Fs numbers. A clear correlation exists between Band 1 and Band 2 centers and Fs number among HED meteorites. Thus, after plotting each meteorite's band centers against its Fs number, we fit a linear relation between the two, yielding the following equations:

\begin{equation}
\textrm{Fs\# (molar \%)} = 664 \times \textrm{B1} - 583 
\label{eq:FsB1}
\end{equation} 

and 

\begin{equation}
\textrm{Fs\# (molar \%)} = 169 \times \textrm{B2} - 296 
\label{eq:FsB2}
\end{equation} 

where B1 and B2 represent the MGM-derived Band 1 and Band 2 centers in microns. These equations are plotted over the Band 1 and Band 2 centers of our HED meteorite sample in Figure \ref{fig:bandvsfs}.

The RMS error of these equations is $\pm$ 6 for Equation \ref{eq:FsB1} and $\pm$ 4 for Equation \ref{eq:FsB2}. This compares favorably to the analogous equations derived by Gaffey et al. (2002) based on their analyses of terrestrial pyroxenes, which have RMS errors of $\pm$ 7 and $\pm$ 4, respectively. Gaffey et al. (2002)'s equations have been used elsewhere to calculate Fs number for V-type asteroids (Burbine et al., 2009; Ieva et al., 2015). The calibration of Fs number is better for Band 2 centers than for Band 1 centers, and so at infinite signal to noise Equation \ref{eq:FsB2} is more accurate. However, telescopic spectra generally display lower S/N around 2 $\mu$m, suggesting that uncertainty in compositional information derived from this band will eventually be dominated by statistical S/N errors as opposed to the systematic RMS errors associated with the absolute calibration of these equations. In circumstances of extremely low S/N in the 2-$\mu$m region, one would want to completely ignore Band 2 and rely solely on Equation \ref{eq:FsB1}. We adopt an intermediate approach of averaging the compositional information derived from Band 1 and Band 2. Fs numbers for each of the meteorites in our sample are shown in Table \ref{tab:hedFs}, along with the Fs numbers predicted by equations (4) and (5) and Gaffey et al. (2002)'s equations.

\subsection{Comparison to non-Vesta-like meteorite spectra} \label{ComparisonMets}

\begin{figure}[H]
\centering
\includegraphics[scale=.5]{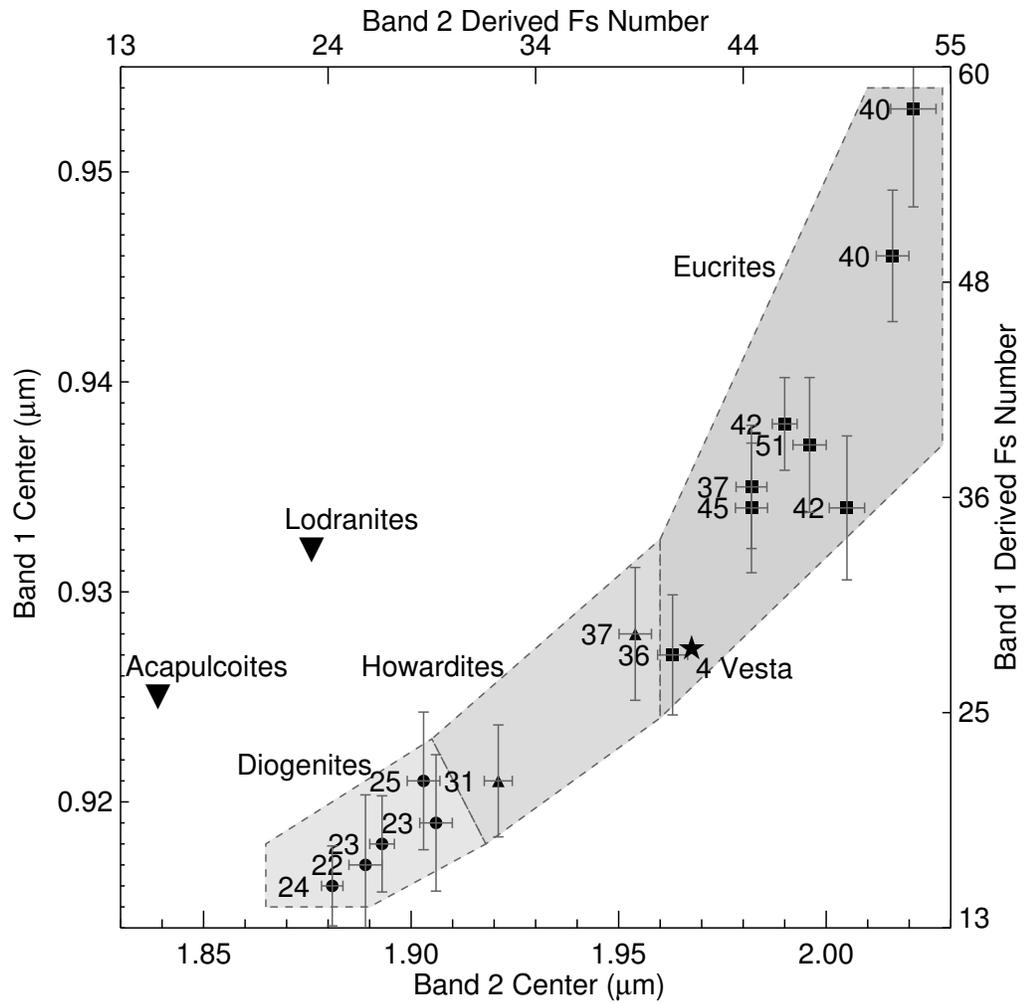}
\caption{Band diagram of HED meteorites. Diogenites are represented by circles, Howardites by triangles, Eucrites by squares, and 4 Vesta by a five-pointed star. The Fs number of each meteorite is shown next to each object. Three ``Fs regions'' are defined, which roughly correspond to the locations of Diogenites, Howardites, and Eucrites. The three regions combined cover the HED "main sequence," the area of band space where most HED meteorites and Vestoids are located. The top and right axes show Fs numbers derived from Band 1 and Band 2 centers using Equations \ref{eq:FsB1} and \ref{eq:FsB2}, respectively.}
\label{fig:fscomparison}
\end{figure}

As these equations were calibrated to HED meteorites with compositions analogous to Vesta, it was important to gauge their accuracy for non-Vesta achondritic compositions. Of particular interest for the non-Vesta basaltic asteroids (where we cannot \textit{a priori} assume a pyroxene-dominated composition like that of Vesta and the HEDs) is the influence of olivine on our derived pyroxene mineralogy. Olivine can artificially increase the MGM-derived Band 1 center, and with it the Fs number derived using Equation \ref{eq:FsB1}. This effect pushes the object above the HED ``Main Sequence''---the region of band parameter space occupied by HED meteorites shown in Figure \ref{fig:fscomparison}. Among our outer belt target objects, at least one---(349) Dembowska---is known to be an olivine-rich body, with a surface olivine:pyroxene ratio perhaps even as high as 1:1 (Gaffey et al., 1993). As such, it was important to test Equations \ref{eq:FsB1} and \ref{eq:FsB2} on olivine-rich meteorites in order to gauge their accuracy.

Acapulcoites and Lodranites are achondrites likely from differentiated or partially differentiated parent bodies and are known to be more olivine-rich than the HED suite. We obtained reflectance spectra of three Acapulcoites and five Lodranites from the RELAB database, all with known compositions, and performed MGM fits on them using the same set of input paramters used for the HED meteorites (see Table \ref{tab:fitparam}). The resulting band centers for each meteorite are shown alongside those of the HEDs in Table \ref{Table:hedMGM}, and the mean band centers for the Acapulcoites and Lodranites are each shown in Figure \ref{fig:fscomparison}.

Acapulcoites and Lodranites were significantly lower in Fs content than their HED counterparts, with Fs numbers of 10.1 molar percent or less. As predicted, the outputs of Equations \ref{eq:FsB1} and \ref{eq:FsB2} do not accurately account for this. Equation \ref{eq:FsB1} is particularly inaccurate, overestimating the Fs numbers of Acapulcoites and Lodranites by averages of 24 and 27, respectively. Equation \ref{eq:FsB2} does a slightly better job, overestimating the Fs numbers of these objects by averages of 8 and 13, respectively. This is to be expected, as the complex olivine features around 1.2 $\mu$m have a significant effect on the apparent Band 1 center, but exert less influence on the apparent Band 2 center.

Since the presence of olivine inflates Band 1 center far more than Band 2 center, the accuracy of our equations can be quantified by examining the ``Fs discrepancy'' between the Fs numbers predicted by each band center. The average Fs discrepancy among our HED meteorite sample was just 2 $\pm$ 2 molar percent, while among our Vestoid sample the average Fs discrepancy was only slightly higher at 4 $\pm$ 3 molar percent. However, the Acapulcoites and Lodranites fell well outside this range, with an average Fs discrepancy of 15 $\pm$ 9. Notably, the Fs discrepancy values for the HEDs and Vestoids are smaller than the RMS error on Equations \ref{eq:FsB1} and \ref{eq:FsB2}, but the Fs discrepancy values for other objects are much larger. Our equations are thus not suitable for analyzing the composition of these objects. However, Figure \ref{fig:fscomparison} demonstrates that our method of MGM fitting is nevertheless capable of clearly differentiating these objects in band space from HED meteorites and Vestoids. Furthermore, our derived pyroxene Fs numbers almost certainly establish an upper limit for these cases.

Two of the outer belt objects listed in Table \ref{tab:astFs} also have Fs discrepancies above the range set by the HED and Vestoid samples: (349) Dembowska (8.5) and (22308) 1990 UO$_4$ (17.7). Our method of compositional analysis thus cannot be used to uniquely determine the Fs content of these objects. However, because they fall above the HED main sequence we are still able to use Equation \ref{eq:FsB1} to set a confident upper limit on their Fs content. This is in keeping with the results of our analysis of Acapulcoites and Lodranites, for which Equations \ref{eq:FsB1} and \ref{eq:FsB2} universally overestimated Fs content.

\begin{table}[H]
\tiny
\centering
\begin{tabular}{| l l l l l l |}
\hline
Object & Fs \# & MGM Calculated B1 Fs \# & MGM Calculated B2 Fs \# & Gaffey B1 Fs \# & Gaffey B2 Fs \# \\
\hline
Diogenites &  &  &  &  & \\																
EETA79002	&	22$^a$	&	26	$\pm$	6	&	23	$\pm$	4	&	25	$\pm$	8	&	28	$\pm$	4	\\
GRO95555	&	25$^a$	&	29	$\pm$	6	&	26	$\pm$	4	&	29	$\pm$	8	&	31	$\pm$	4	\\
Johnstown	&	24$^b$	&	25	$\pm$	6	&	22	$\pm$	4	&	24	$\pm$	7	&	27	$\pm$	4	\\
LAP91900		&	23$^a$	&	27	$\pm$	6	&	26	$\pm$	4	&	27	$\pm$	8	&	32	$\pm$	4	\\
Tatahouine	&	23$^b$	&	27	$\pm$	6	&	24	$\pm$	4	&	26	$\pm$	7	&	29	$\pm$	4	\\
 &  &  &  &  & \\
Eucrites &  &  &  &  & \\
Chervony Kut	&	42$^a$	&	37	$\pm$	6	&	43	$\pm$	4	&	42	$\pm$	8	&	52	$\pm$	4	\\
EET87520	&	40$^a$	&	45	$\pm$	6	&	45	$\pm$	4	&	54	$\pm$	8	&	55	$\pm$	4	\\
GRA98098	&	51$^a$	&	39	$\pm$	6	&	41	$\pm$	4	&	45	$\pm$	8	&	50	$\pm$	4	\\
Ibitira		&	42$^a$	&	39	$\pm$	6	&	40	$\pm$	4	&	46	$\pm$	7	&	49	$\pm$	4	\\
MET01081	&	45$^a$	&	37	$\pm$	6	&	39	$\pm$	4	&	42	$\pm$	8	&	48	$\pm$	4	\\
Moore County	&	37$^a$	&	37	$\pm$	6	&	39	$\pm$	4	&	43	$\pm$	8	&	48	$\pm$	4	\\
PCA91078	&	40$^a$	&	49	$\pm$	7	&	46	$\pm$	4	&	62	$\pm$	8	&	56	$\pm$	4	\\
Serra de Mage	&	36$^a$	&	32	$\pm$	6	&	36	$\pm$	4	&	35	$\pm$	8	&	44	$\pm$	4	\\
 &  &  &  &  & \\
Howardites &  &  &  &  & \\
EET87503	&	37$^a$	&	33	$\pm$	6	&	34	$\pm$	4	&	36	$\pm$	8	&	42	$\pm$	4	\\
QUE94200	&	31$^a$	&	29	$\pm$	6	&	29	$\pm$	4	&	29	$\pm$	8	&	35	$\pm$	4	\\
\hline
$^a$Mayne et al. (2011) & & & & &\\
$^b$Burbine et al. (2009)& & & & &\\
\hline
\end{tabular}
\caption{Measured and predicted Fs numbers for HED meteorites. In all cases, systematic error on the equations used to determine Fs number (see section \ref{Composition}) dominates over errors in determining band center. Both the MGM-derived and Gaffey et a. (2002) equations are sufficient to confine a given object to within one of the Fs regions shown in Figure \ref{fig:fscomparison}, provided it falls on or near the HED ``main sequence'' on the band diagram.}\label{tab:hedFs}
\end{table}

\begin{table}[H]
\tiny
\centering
\begin{tabular}{| l l l l l l l |}
\hline
\multicolumn{7}{ | c | }{Vesta and Vestoids}\\
\hline
Object & Albedo & Temp. (K) & B1 Corr. ($\mu$m) & B2 Corr. ($\mu$m) & B1 Center ($\mu$m) & B2 Center ($\mu$m) \\
\hline
(4) Vesta	&	0.42	&	161	&	0.002	&	0.025	&	0.927	$\pm$	0.002	&	1.968	$\pm$	0.003	\\
(809) Lundia	&	0.33	&	186	&	0.002	&	0.020	&	0.927	$\pm$	0.002	&	1.965	$\pm$	0.006	\\
(956) Elisa	&	0.15	&	202	&	0.002	&	0.017	&	0.923	$\pm$	0.001	&	1.941	$\pm$	0.005	\\
(1468) Zomba	&	0.40	&	199	&	0.002	&	0.018	&	0.927	$\pm$	0.002	&	1.977	$\pm$	0.005	\\
(1929) Kollaa	&	0.39	&	170	&	0.002	&	0.023	&	0.934	$\pm$	0.002	&	1.967	$\pm$	0.005	\\
(2045) Peking	&	0.25	&	168	&	0.002	&	0.023	&	0.933	$\pm$	0.002	&	1.970	$\pm$	0.007	\\
(2371) Dimitrov	&	0.35	&	178	&	0.002	&	0.024	&	0.926	$\pm$	0.003	&	1.969	$\pm$	0.024	\\
(2442) Corbett	&	0.40	&	168	&	0.002	&	0.023	&	0.925	$\pm$	0.002	&	1.951	$\pm$	0.007	\\
(2511) Patterson	&	0.29	&	176	&	0.002	&	0.022	&	0.927	$\pm$	0.002	&	1.957	$\pm$	0.005	\\
(2566) Kirghizia		&	0.26	&	171	&	0.002	&	0.023	&	0.930	$\pm$	0.002	&	1.958	$\pm$	0.005	\\
(2579) Spartacus	&	0.53	&	159	&	0.002	&	0.025	&	0.921	$\pm$	0.000	&	1.997	$\pm$	0.014	\\
(2653) Principia	&	0.40	&	158	&	0.002	&	0.025	&	0.925	$\pm$	0.002	&	1.980	$\pm$	0.004	\\
(2763) Jeans	&	0.41	&	167	&	0.002	&	0.024	&	0.933	$\pm$	0.002	&	2.003	$\pm$	0.003	\\
(2795) Lepage	&	0.40	&	167	&	0.002	&	0.024	&	0.926	$\pm$	0.002	&	1.965	$\pm$	0.003	\\
(2823) van der Laan	&	0.32	&	175	&	0.002	&	0.022	&	0.923	$\pm$	0.002	&	1.964	$\pm$	0.027	\\
(2851) Harbin	&	0.40	&	162	&	0.002	&	0.025	&	0.917	$\pm$	0.002	&	1.929	$\pm$	0.004	\\
(2912) Lapalma	&	0.40	&	172	&	0.002	&	0.023	&	0.921	$\pm$	0.002	&	1.942	$\pm$	0.005	\\
(3155) Lee	&	0.40	&	157	&	0.002	&	0.025	&	0.920	$\pm$	0.002	&	1.914	$\pm$	0.008	\\
(3344) Modena	&	0.40	&	171	&	0.002	&	0.023	&	0.920	$\pm$	0.002	&	1.909	$\pm$	0.006	\\
(3657) Ermolova	&	0.40	&	170	&	0.002	&	0.023	&	0.923	$\pm$	0.002	&	1.948	$\pm$	0.014	\\
(3703) Volkonskaya	&	0.24	&	178	&	0.002	&	0.022	&	0.917	$\pm$	0.002	&	1.953	$\pm$	0.009	\\
(3782) Celle	&	0.50	&	155	&	0.002	&	0.026	&	0.927	$\pm$	0.002	&	1.947	$\pm$	0.005	\\
(4038) Kristina	&	0.40	&	174	&	0.002	&	0.022	&	0.911	$\pm$	0.001	&	1.965	$\pm$	0.006	\\
(4188) Kitezh	&	0.34	&	178	&	0.002	&	0.022	&	0.935	$\pm$	0.002	&	1.965	$\pm$	0.004	\\
(4215) Kamo	&	0.40	&	158	&	0.002	&	0.025	&	0.920	$\pm$	0.002	&	1.952	$\pm$	0.008	\\
(4796) Lewis	&	0.40	&	160	&	0.002	&	0.025	&	0.929	$\pm$	0.001	&	1.948	$\pm$	0.018	\\
(5481) Kiuchi	&	0.40	&	163	&	0.002	&	0.024	&	0.923	$\pm$	0.002	&	1.949	$\pm$	0.011	\\
(5498) Gustafsson	&	0.23	&	193	&	0.002	&	0.019	&	0.936	$\pm$	0.001	&	1.981	$\pm$	0.005	\\
(7800) Zhongkeyuan	&	0.39	&	159	&	0.002	&	0.025	&	0.923	$\pm$	0.000	&	1.918	$\pm$	0.032	\\
(9481) Menchu	&	0.20	&	171	&	0.002	&	0.023	&	0.929	$\pm$	0.000	&	1.939	$\pm$	0.025	\\
(9553) Colas	&	0.18	&	193	&	0.002	&	0.019	&	0.923	$\pm$	0.002	&	1.929	$\pm$	0.008	\\
(16416) 1987 SM3	&	0.47	&	158	&	0.002	&	0.025	&	0.922	$\pm$	0.001	&	1.971	$\pm$	0.020	\\
(26886) 1994 TJ2	&	0.24	&	185	&	0.002	&	0.020	&	0.918	$\pm$	0.003	&	1.917	$\pm$	0.025	\\
(27343) 2000 CT102	&	0.34	&	185	&	0.002	&	0.020	&	0.921	$\pm$	0.002	&	1.926	$\pm$	0.012	\\
(33881) 2000 JK66	&	0.53	&	178	&	0.002	&	0.022	&	0.923	$\pm$	0.002	&	1.955	$\pm$	0.013	\\
(36412) 2000 OP49	&	0.37	&	175	&	0.002	&	0.022	&	0.927	$\pm$	0.001	&	1.971	$\pm$	0.020	\\
(38070) 1999 GG2	&	0.40	&	182	&	0.002	&	0.021	&	0.932	$\pm$	0.002	&	1.963	$\pm$	0.016	\\
(50098) 2000 AG98	&	0.36	&	183	&	0.002	&	0.021	&	0.926	$\pm$	0.002	&	1.961	$\pm$	0.005	\\
(97276) 1999 XC143		&	0.40	&	176	&	0.002	&	0.022	&	0.933	$\pm$	0.003	&	2.013	$\pm$	0.013	\\
\hline
\multicolumn{7}{ | c | }{Outer Belt Objects}\\
\hline
(349) Dembowska	&	0.14	&	160	&	0.002	&	0.025	&	0.931	$\pm$	0.001	&	1.911	$\pm$	0.001	\\
(1459) Magnya	&	0.22	&	137	&	0.003	&	0.029	&	0.933	$\pm$	0.000	&	1.954	$\pm$	0.000	\\
(21238) Panarea	&	0.37	&	187	&	0.002	&	0.022	&	0.929	$\pm$	0.000	&	1.934	$\pm$	0.005	\\
(22308) 1990 UO$_4$	&	0.16	&	171	&	0.002	&	0.023	&	0.930	$\pm$	0.000	&	1.852	$\pm$	0.001	\\
(40521) 1999 RL$_{95}$	&	0.28	&	221	&	0.001	&	0.014	&	0.931	$\pm$	0.000	&	1.941	$\pm$	0.003	\\
(105041) 2000 KO$_{41}$	&	0.40	&	148	&	0.003	&	0.027	&	0.907	$\pm$	0.000	&	1.866	$\pm$	0.002	\\
\hline
\multicolumn{7}{ | c | }{Unclassified Objects (See Section \ref{NewType})}\\
\hline
(7472) Kumakiri	&	0.28	&	144	&	0.003	&	0.028	&	1.015	$\pm$	0.001	&	1.963	$\pm$	0.000	\\
(10537) 1991 RY$_{16}$	&	0.31	&	154	&	0.002	&	0.026	&	0.952	$\pm$	0.000	&	1.931	$\pm$	0.001	\\
(14390) 1990 QP$_{19}$	&	0.22	&	158	&	0.002	&	0.025	&	0.973	$\pm$	0.001	&	1.994	$\pm$	0.000	\\
\hline
\end{tabular}
\caption{MGM fit results for main belt basaltic asteroids. The columns in this table are: object designation, albedo, estimated mean surface temperature (see Equation \ref{eq:tempeq}), temperature correction to B1 center, temperature correction to B2 center, temperature-corrected B1 center, temperature correction to B2 center.}
\label{tab:astMGM}
\end{table}

\section{Discussion} \label{Discussion}

\subsection{Spectral Outliers} \label{NewType}

Three objects initially classified as V-types based on visible wavelength spectra ((7472) Kumakiri, (14390) 1990 QP$_{19}$, and (10537) 1991 RY$_{16}$) are found to be spectrally unique from the remainder of our sample due to a wide and complex 1-$\mu$m absorption band and a wide and shallow 2-$\mu$m absorption band. The shape of the 1-$\mu$m band is likely due to overlap with an unusually large absorption feature or features in the 1.2-$\mu$m region. This is particularly visible in spectra of (14390) 1990 QP$_{19}$ and (10537) 1991 RY$_{16}$, both of which show a distinct rounded shoulder at the upper end of their 1-$\mu$m bands---a feature which is far less obvious in the spectrum of asteroid (7472) Kumakiri. 

The shallowness of these objects' 2-$\mu$m absorption band results in a very different ratio of band depths between the 1- and 2-$\mu$m bands when compared to typical V-type asteroids. For most V-types, the depth of the 1-$\mu$m band is typically between 1 and 1.5 times the depth of the 2-$\mu$m band. However, in the case of all three of these objects, that ratio is significantly higher: 2.25 ((10537) 1991 RY$_{16}$), 2.21 ((14390) 1990 QP$_{19}$) and, in a much more extreme case, 10.74 (Kumakiri). The Bus-DeMeo taxonomy utility run by the Planetary Spectroscopy Group at MIT (DeMeo et al. 2009) classified (10537) 1991 RY$_{16}$ as a V-type, though the fit was poor, and was unable to assign a spectral type to (7472) Kumakiri or (14390) 1990 QP$_{19}$, though Burbine el al. (2011) suggested that (7472) Kumakiri might be a member of the O-type. As shown in Figure \ref{fig:outliers}, (10537) 1991 RY$_{16}$'s 2-$\mu$m band is far shallower than that of a typical V-type, and neither (7472) Kumakiri nor (14390) 1990 QP$_{19}$ is an exact match to the only spectroscopically confirmed O-type asteroid, 3628 Boznemcova.

\begin{figure}[H]
\centering
\includegraphics[scale=.55]{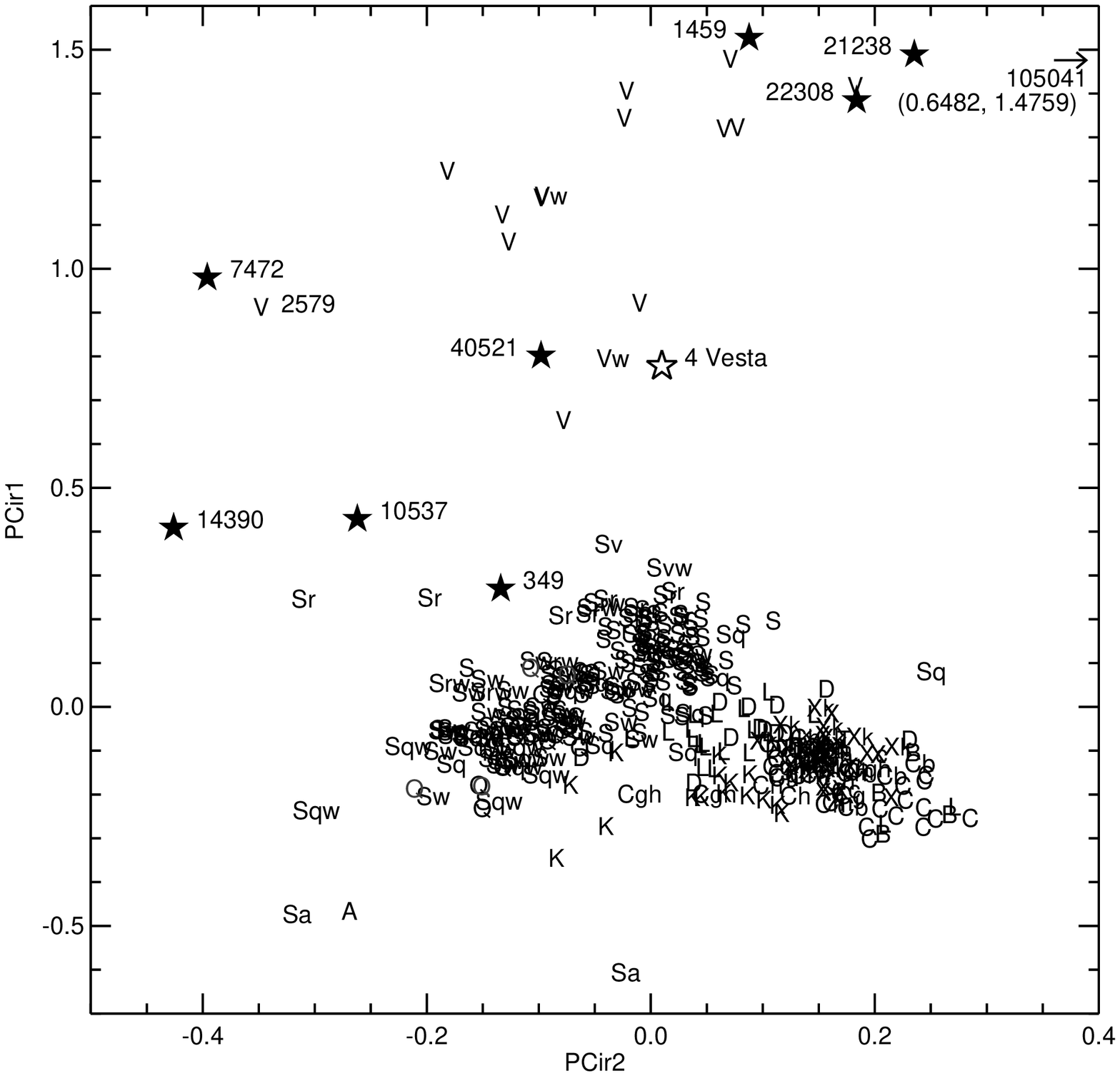}
\caption{Principal component diagram showing major asteroid types. Objects examined in this study are represented by five-pointed stars, with the exception of outer belt V-type (105041) 2000 KO$_{41}$ (located outside the plot range at PCir2 = 0.6482, PCir1 = 1.4759). Kumakiri, Magnya, and (10537) 1991 RY$_{16}$ fall in a region in PC space which is otherwise entirely empty save for the unusual V-type (2579) Spartacus. The only known O-type, 3628 Boznemcova, is also located off the plot at PCir2 = -0.6355, PCir1 = -0.1751.}
\label{fig:pcdiagram}
\end{figure}

The distinctiveness of these objects is further underscored by where they fall in principal component space. Principal Component Analysis (PCA) reduces the dimensionality of data, using coordinate transformations to minimize variance along two axes: Principal Component 1 and Principal Component 2 (DeMeo et al., 2009). PCA is described in detail in Tholen (1984) and Bus (1999). Figure \ref{fig:pcdiagram} shows that the three objects are located in an almost entirely empty region in the range PC2 = -0.2 to -0.5, and PC1 = 0.3 to 1.0 (these values, and those shown in Figure \ref{fig:pcdiagram}, refer to NIR principal component values, which are distinct from their visible or visible + NIR counterparts). The only other object inhabiting this region is the V-type asteroid (2579) Spartacus, which is a known oddity among Vestoids (Burbine et al., 2001). Spartacus is known to have a much more complex 1-$\mu$m band than most V-types, and exhibits an unusually large Fs discrepancy for a Vestoid: Equations \ref{eq:FsB1} and \ref{eq:FsB2} predict Fs numbers of 28.8 and 41.5, respectively (straddling the range of Fs numbers that would be expected from a Howardite). However, neither it nor the Acapulcoites and Lodranites share any of the gross morphological features---in particular, the extremely shallow 2-$\mu$m band---which distinguish (10537) 1991 RY$_{16}$, (14390) 1990 QP$_{19}$, and (7472) Kumakiri from other spectral types. Based on these three asteroids' location in this sparse region of principal component space, they may require a significant expansion in principal component space of the V-type complex or the creation of a new spectral type.

In previous examinations of Spartacus, it has been speculated that its complex 1-$\mu$m feature may be due either to compositional variation across the surface of the asteroid or to a much higher concentration of olivine than is present on the surface of other Vestoids (Burbine et al., 2001). While significant variation in surface composition is uncommon among asteroids, the three overlapping olivine bands known to fall in the 1-$\mu$m region are a plausible explanation for this feature of Spartacus' spectrum and the spectra presented here. However, while its large Fs discrepancy is reminiscent of the Acapulcoites and Lodranites analyzed in this work, the fact that Equation \ref{eq:FsB2} predicts a higher Fs number than Equation \ref{eq:FsB1} is not what would be expected from olivine features in its spectrum. Furthermore, distinguishing between the three overlapping olivine bands and the two overlapping pyroxene bands in the 1-$\mu$m region is extremely challenging, and so we draw no conclusions regarding the specific mineralogy of these unusual objects.

\subsection{Band centers and Fs numbers for outer belt objects} \label{FsNums}

\begin{figure}[H]
\centering
\includegraphics[scale=.4545]{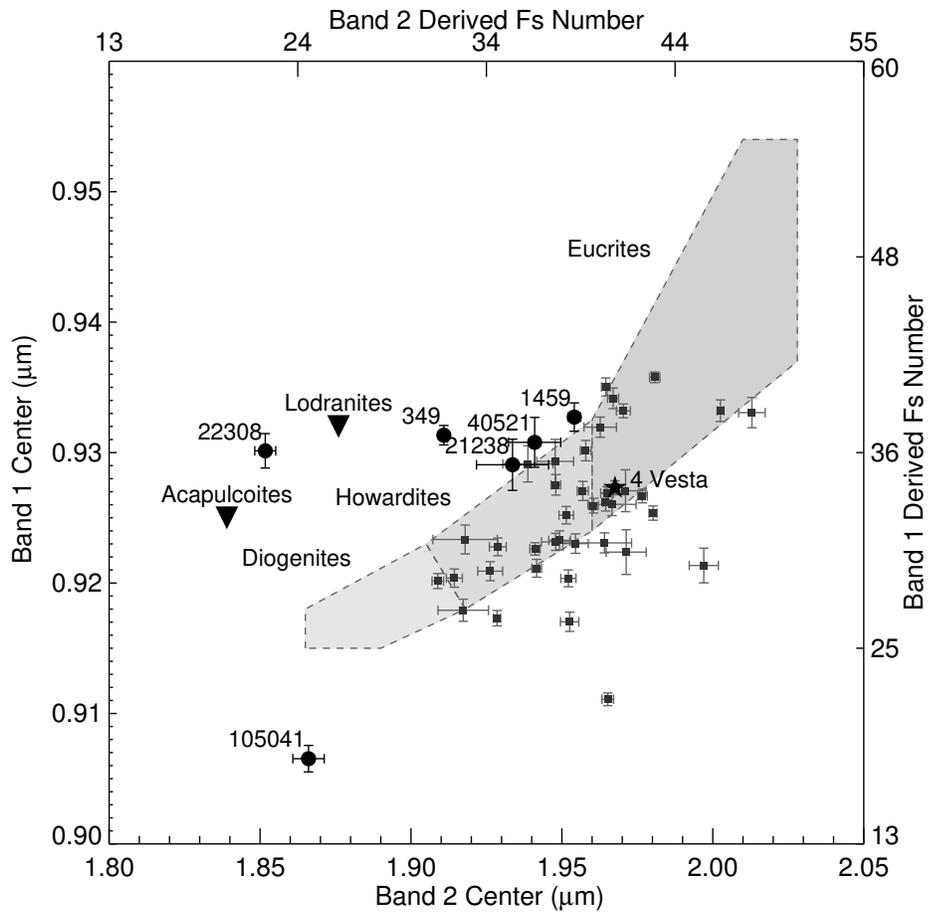}
\caption{Band diagram showing Vestoids and outer belt basaltic asteroids plotted over Diogenite, Howardite, and Eucrite regions. Vesta is represented as a five-pointed star, main-belt Vestoids as squares, and outer belt basaltic asteroids as filled circles. The average band centers of comparison Lodranites and Acapulcoites are shown as filled triangles.}
\label{fig:bandplot}
\end{figure}

As shown in Figure \ref{fig:bandplot}, the six outer belt V- and R-type objects all fall outside of the HED main sequence, and are primarily offset to the left relative to Vesta and the inner belt Vestoids. This supports the results of Ieva et al. (2015), who used MGM to locate (1459) Magnya and (21238) Panarea well outside the region of band space occupied by the Vestoid family and HED meteorites. This distinction is further underlined by comparing a weighted mean of Band 1 and Band 2 centers for each group of objects: the mean band centers for the six outer belt V- and R-types, weighted by signal to noise, are 0.930 $\pm$ 0.025 and 1.918 $\pm$ 0.049, as opposed to 0.926 $\pm$ 0.013 and 1.96 $\pm$ 0.024 for Vesta and the Vestoids. This comparison also serves to further differentiate the three anomalous asteroids 14290, 10537, and (7472) Kumakiri, whose weighted mean band centers are much higher (particularly in Band 1) at 0.980 $\pm$ 0.032 and 1.971 $\pm$ 0.032 . Performing a Kolmogorov-Smirnov test on the distribution of Vestoid and outer belt V- and R-type Band 1 and 2 centers showed that the two populations are different in Band 1 center at 2.1 $\sigma$ confidence and in Band 2 center at 2.35 $\sigma$ confidence.

After fitting the spectra of our outer belt asteroids using MGM, we calculated their predicted Fs numbers with Equations \ref{eq:FsB1} and \ref{eq:FsB2}. We determined error on these objects' Fs numbers by propagating the error on their band centers through Equations \ref{eq:FsB1} and \ref{eq:FsB2} and combining the result in quadrature with the RMS error on the equations themselves. Performing the same process on the weighted mean band centers calculated above allowed for mineralogical comparisons between these groups of objects. However, as shown in Table \ref{tab:astFs} and Figure \ref{fig:bandplot}, the farther from Vesta and the Vestoids an object lies in the Band 1 vs Band 2 plane, the higher the discrepancy is between the Fs numbers predicted by Equations \ref{eq:FsB1} and \ref{eq:FsB2}: (1459) Magnya sits relatively close to Vesta, and has a discrepancy of only 2 molar percent; however, (22308) Panarea is well to the left of Vesta and the entire Vestoid population, and has a discrepancy of 18 molar percent. Using this discrepancy as a proxy for the accuracy of these equations shows that these equations do not accurately estimate the Fs number of (7472) Kumakiri, (10537) 1991 RY$_{16}$, and (14390) 1990 QP$_{19}$: in the most extreme case, that of (7472) Kumakiri, the values predicted by the two equations differ by 65 molar percent. 

Fortunately, for most of our outer belt V- and R-types, these equations yield consistent results and thus can constrain each object's Fs number to a relatively narrow range, usually less than ten molar percent. The two exceptions to this, as discussed in section \ref{ComparisonMets}, are (349) Dembowska and (22308) 1990 UO$_4$. The former of these is known to be an olivine-rich body. The latter's location in band space is extremely close to the area inhabited by the Acapulcoites, and thus may have an Acapulcoite-like composition as well, which would imply the presence of olivine and a low Fs number. As a result, a meaningful comparison can be made between the Vestoids and most of our outer belt V-types using the average band centers of each, and the two objects which are not easily compared are nevertheless likely to be significantly lower in Fs content than the Vestoids.

The weighted mean band centers for the Vestoids yield predicted Fs numbers of 32 $\pm$ 9 and 35 $\pm$ 4. This range is similar to a Howardite or low-Fs Eucrite meteorite (see Figure \ref{fig:fscomparison}), which is to be expected given that Vestoids are comprised primarily of material from Vesta's crust and upper mantle. These values are lower for our outer belt targets due to their significantly lower average Band 2 center. For these six objects, Equations \ref{eq:FsB1} and \ref{eq:FsB2} predict average Fs numbers of 35 $\pm$ 17 and 28 $\pm$ 8, respectively. While there is overlap between the Fs numbers predicted for the outer belt V- and R-types and the Vestoids, these results suggest that the typical Fs content of these objects may be lower than their inner-belt Vestoid counterparts. This is particularly true since the actual difference in Fs content may well be more significant than indicated by raw comparison of these numbers due to the possible presence of olivine artificially raising the Fs numbers of (349) Dembowska and (22308) 1990 UO$_4$. However, these results are somewhat speculative and based on small sample sizes. As we discuss in section \ref{Formation}, more data are needed in order to further constrain the composition of other igneous asteroids in the outer main belt, which could in turn provide insights regarding the environments in which they formed.

\begin{table}[H]
\tiny
\centering
\begin{tabular}{| l l l l l |}
\hline
\multicolumn{5}{ | c | }{Vesta and Vestoids}\\
\hline
Object & MGM Predicted B1 Fs & MGM Predicted B2 Fs & Gaffey B1 Fs & Gaffey B2 Fs\\
\hline
(4) Vesta	&	33	$\pm$	6	&	37	$\pm$	4	&	35	$\pm$	7	&	44	$\pm$	4	\\
(956) Elisa	&	30	$\pm$	6	&	32	$\pm$	4	&	30	$\pm$	7	&	39	$\pm$	4	\\
(1468) Zomba	&	32	$\pm$	6	&	38	$\pm$	4	&	35	$\pm$	7	&	46	$\pm$	4	\\
(1929) Kollaa	&	37	$\pm$	6	&	36	$\pm$	4	&	42	$\pm$	7	&	44	$\pm$	4	\\
(2045) Peking	&	37	$\pm$	6	&	37	$\pm$	4	&	41	$\pm$	7	&	45	$\pm$	4	\\
(2371) Dimitrov	&	32	$\pm$	6	&	36	$\pm$	4	&	34	$\pm$	7	&	44	$\pm$	4	\\
(2442) Corbett	&	31	$\pm$	6	&	34	$\pm$	4	&	33	$\pm$	7	&	41	$\pm$	4	\\
(2511) Patterson	&	33	$\pm$	6	&	35	$\pm$	4	&	35	$\pm$	7	&	42	$\pm$	4	\\
(2566) Kirghizia	&	35	$\pm$	6	&	35	$\pm$	4	&	38	$\pm$	7	&	43	$\pm$	4	\\
(2579) Spartacus	&	29	$\pm$	6	&	41	$\pm$	4	&	29	$\pm$	7	&	51	$\pm$	4	\\
(2653) Principia	&	31	$\pm$	6	&	39	$\pm$	4	&	33	$\pm$	7	&	47	$\pm$	4	\\
(2763) Jeans	&	37	$\pm$	6	&	42	$\pm$	4	&	41	$\pm$	7	&	52	$\pm$	4	\\
(2795) Lepage	&	32	$\pm$	6	&	36	$\pm$	4	&	34	$\pm$	7	&	44	$\pm$	4	\\
(2823) van der Laan	&	30	$\pm$	6	&	36	$\pm$	4	&	31	$\pm$	7	&	44	$\pm$	4	\\
(2851) Harbin	&	26	$\pm$	6	&	30	$\pm$	4	&	25	$\pm$	7	&	36	$\pm$	4	\\
(2912) Lapalma	&	29	$\pm$	6	&	32	$\pm$	4	&	29	$\pm$	7	&	39	$\pm$	4	\\
(3155) Lee	&	28	$\pm$	6	&	28	$\pm$	4	&	28	$\pm$	7	&	34	$\pm$	4	\\
(3344) Modena	&	28	$\pm$	6	&	27	$\pm$	4	&	28	$\pm$	7	&	32	$\pm$	4	\\
(3657) Ermolova	&	30	$\pm$	6	&	33	$\pm$	4	&	31	$\pm$	7	&	40	$\pm$	4	\\
(3703) Volkonskaya	&	26	$\pm$	6	&	34	$\pm$	4	&	25	$\pm$	7	&	41	$\pm$	4	\\
(3782) Celle	&	33	$\pm$	6	&	33	$\pm$	4	&	35	$\pm$	7	&	40	$\pm$	4	\\
(4038) Kristina	&	22	$\pm$	6	&	36	$\pm$	4	&	19	$\pm$	7	&	44	$\pm$	4	\\
(4188) Kitezh	&	38	$\pm$	6	&	36	$\pm$	4	&	43	$\pm$	7	&	44	$\pm$	4	\\
(4215) Kamo	&	28	$\pm$	6	&	34	$\pm$	4	&	28	$\pm$	7	&	41	$\pm$	4	\\
(4796) Lewis	&	34	$\pm$	6	&	33	$\pm$	4	&	37	$\pm$	7	&	40	$\pm$	4	\\
(5481) Kiuchi	&	30	$\pm$	6	&	33	$\pm$	4	&	31	$\pm$	7	&	41	$\pm$	4	\\
(5498) Gustafsson	&	38	$\pm$	6	&	39	$\pm$	4	&	44	$\pm$	7	&	47	$\pm$	4	\\
(7800) Zhongkeyuan	&	30	$\pm$	6	&	28	$\pm$	4	&	31	$\pm$	7	&	34	$\pm$	4	\\
(809) Lundia	&	32	$\pm$	6	&	36	$\pm$	4	&	35	$\pm$	7	&	44	$\pm$	4	\\
(9481) Menchu	&	34	$\pm$	6	&	32	$\pm$	4	&	37	$\pm$	7	&	39	$\pm$	4	\\
(9553) Colas	&	30	$\pm$	6	&	30	$\pm$	4	&	31	$\pm$	7	&	37	$\pm$	4	\\
(16416) 1987 SM$_3$	&	29	$\pm$	6	&	37	$\pm$	4	&	30	$\pm$	7	&	45	$\pm$	4	\\
(26886) 1994 TJ$_2$	&	26	$\pm$	6	&	28	$\pm$	4	&	26	$\pm$	7	&	34	$\pm$	4	\\
(27343) (2000 CT$_{102}$	&	28	$\pm$	6	&	30	$\pm$	4	&	29	$\pm$	7	&	36	$\pm$	4	\\
(33881) 2000 JK$_{66}$	&	30	$\pm$	6	&	34	$\pm$	4	&	31	$\pm$	7	&	42	$\pm$	4	\\
(36412) 2000 OP$_{49}$	&	33	$\pm$	6	&	37	$\pm$	4	&	35	$\pm$	7	&	45	$\pm$	4	\\
(38070) 1999 GG$_2$	&	36	$\pm$	6	&	36	$\pm$	4	&	40	$\pm$	7	&	43	$\pm$	4	\\
(50098) 2000AG$_{98}$	&	32	$\pm$	6	&	35	$\pm$	4	&	34	$\pm$	7	&	43	$\pm$	4	\\
(97276) 1999XC$_{143}$	&	37	$\pm$	7	&	44	$\pm$	4	&	41	$\pm$	8	&	54	$\pm$	4	\\
\hline
\multicolumn{5}{ | c | }{Outer belt objects}\\
\hline                
(349) Dembowska	&	35	$\pm$	6	&	27	$\pm$	4	&	39	$\pm$	7	&	33	$\pm$	4	\\
(1459) Magnya	&	36	$\pm$	6	&	34	$\pm$	4	&	41	$\pm$	7	&	42	$\pm$	4	\\
(21238 Panarea	&	35	$\pm$	6	&	32	$\pm$	4	&	40	$\pm$	7	&	37	$\pm$	4	\\
(22308) 1990 UO$_4$	&	35	$\pm$	6	&	17	$\pm$	4	&	38	$\pm$	7	&	21	$\pm$	4	\\
(40521) 1999 RL$_5$	&	35	$\pm$	6	&	25	$\pm$	4	&	39	$\pm$	7	&	29	$\pm$	4	\\
(105041) 2000 KO$_{41}$	&	19	$\pm$	6	&	19	$\pm$	4	&	14	$\pm$	7	&	24	$\pm$	4	\\
\hline
\multicolumn{5}{| c |}{Unclassified objects (See Section \ref{NewType})}\\
\hline             
(7472) Kumakiri	&	91	$\pm$	7	&	36	$\pm$	4	&	125	$\pm$	7	&	44	$\pm$	4	\\
(10537) 1991 RY$_{16}$	&	49	$\pm$	6	&	30	$\pm$	4	&	61	$\pm$	7	&	37	$\pm$	4	\\
(14390) 1990 QP$_{19}$	&	63	$\pm$	7	&	41	$\pm$	4	&	82	$\pm$	7	&	50	$\pm$	4	\\
\hline
\end{tabular}
\caption{Fs numbers for asteroids calculated from MGM-derived equations and Gaffey et al. (2002)'s equations. In nearly all cases, error is dominated by systematic uncertainty from Equations \ref{eq:FsB1} and \ref{eq:FsB2}. The outputs of these equations correspond very closely to one another for Vesta and the Vestoids; however, the discrepancy between the predicted Fs number from each equation grows with the object's distance from the HED main sequence.}
\label{tab:astFs}
\end{table}

\subsection{Compositional implications for formation} \label{Formation}

These asteroids' lower average Fs numbers are of interest due to ferrosilite's ability to act as a tracer of the oxidation state of a mineral's formation environment. In oxidizing conditions, iron is preferentially incorporated into silicate minerals as they form, while reducing environments tend to send more iron into the metallic state. In the case of asteroids, this means that objects composed of high-Fs-number pyroxene are likely to have formed in a more oxidizing environment than their low-Fs-number counterparts. That the average Fs number of our outer belt targets may be, on average, somewhat lower than the Vestoids' is indicative of these objects having formed in a more reducing environment.

This fact is of particular significance when we consider it in light of the oxidation gradient that existed within the solar nebula. Due to the increased presence of water at lower temperatures, a newly-forming planetesimal was subject to significantly more oxidizing conditions if it formed farther away from the sun. Currently, our target objects orbit well beyond Vesta---in some cases, nearly one full astronomical unit farther from the Sun. However, they show much more dispersion in parameter space than the Vestoids, generally featuring lower Fs numbers and greater diversity in Fs number. This then implies that, despite their current orbits, at least some of these objects' progenitor planetesimals may have formed in other regions of the solar nebula or at different times than Vesta. If this is indeed the case, it may support the Bottke et al. (2006) model of scattering in the young solar system, which suggests that numerous differentiated asteroids formed interior to the main belt and were later scattered into stable main belt orbits. Alternatively, it may be evidence of the ``Grand Tack'' hypothesis of Walsh et al. (2012), in which Jupiter's inward and subsequent outward migration scatters material into the asteroid belt from two sources---one inside Jupiter's formation region at approximately 3.5 AU, and one between Jupiter and Saturn. These observational results may be of use in establishing constraints on both of these models.

\section{Summary} \label{Summary}

We present visible and near-infrared spectral analysis of nine outer belt basaltic asteroids, 47 Vestoids, fifteen HED meteorites, and eight Acapulcoite and Lodranite meteorites. MGM band fitting allowed us to extract precise information about the 1- and 2-$\mu$m absorption features in each spectrum. Our procedure for analyzing error in MGM-derived band centers is discussed for use in future works. Three outer belt objects had band centers and principal component values which were inconsistent with any currently defined taxonomic classification. We correlated the band centers of HED meteorites with known composition to their Fs numbers, then applied those relationships to the spectra of Vestoids and outer belt basaltic objects to determine their bulk pyroxene Fs content. Acapulcoite and Lodranite spectra were used as comparators to evaluate the accuracy of this technique for non-HED but still igneous compositions. Average Fs numbers of outer belt objects were, on average, slightly lower than the Fs numbers of Vestoids. Due to the gradient in oxidation state which existed in the solar nebula, this may suggest that these objects formed in a different location or time relative to Vesta. This result may support either the Bottke et al. (2006) model of planetesimal scattering in the early solar system or the Walsh et al. (2012) model of scattering due to giant planet migration. However, these conclusions are based on small sample sizes, and additional work is required to better understand compositions of these objects and the specific oxidation state of the solar nebula as a function of time and location.

\section{Acknowledgements} \label{Acknowledgements}

This work was partially supported by the NSF under its Research Experience for Undergraduates (REU) program at Northern Arizona University and Lowell Observatory. T. L. acknowledges the hard and skillful work of Kathy Eastwood as director of that program, the invaluable support of all of its participants, and the logistical assistance provided by Kathleen Stigmon and Natalie Sheehan. This work made use of meteorite reflectance spectra obtained by several investigators at the NASA RELAB facility at Brown University, without which this study would never have been possible. Much of the data utilized in this publication were obtained and made available by the The MIT-UH-IRTF Joint Campaign for NEO Reconnaissance. The IRTF is operated by the University of Hawaii under Cooperative Agreement no. NCC 5-538 with the National Aeronautics and Space Administration, Office of Space Science, Planetary Astronomy Program. The MIT component of this work is supported by NASA grant 09-NEOO009-0001, and by the National Science Foundation under Grants Nos. 0506716 and 0907766. N.M acknowledges support from an NSF Astronomy and Astrophysics Postdoctoral Fellowship (2012-2015) and Lowell Observatory. Acquisition of data from FIRE at Las Campanas Observatory were supported by a postdoctoral fellowship to N.M. by the Carnegie Institute of Washington Department of Terrestrial Magnetism. F.E.D. acknowledges support by the National Aeronautics and Space Administration under Grant No. NNX12AL26G issued through the Planetary Astronomy Program. Much of the data used in this study was taken using telescopes located at the summit of Mauna Kea, which holds enormous cultural and spiritual significance to the native Hawaiian community. We are immensely grateful for the opportunity to make use of this important site for astronomical observations.

\end{document}